# Data Science in Olfaction

**Vivek K Agarwal**[1,2], **Joshua S. Harvey**[2], **Dmitry Rinberg**[2,3,4], **Vasant Dhar**[1,5]

[1]Center for Data Science, New York University
[2]Neuroscience Institute, NYU Langone Health
[3]Center for Neural Science, New York University
[4]Department of Physics, New York University
[5]Stern School of Business, New York University

**Abstract**

Advances in neural sensing technology are making it possible to observe the olfactory process in great detail. In this paper, we conceptualize smell from a Data Science and AI perspective, that relates the properties of odorants to how they are sensed and analyzed in the olfactory system from the nose to the brain. Drawing distinctions to color vision, we argue that smell presents unique measurement challenges, including the complexity of stimuli, the high dimensionality of the sensory apparatus, as well as what constitutes ground truth. In the face of these challenges, we argue for the centrality of odorant-receptor interactions in developing a theory of olfaction. Such a theory is likely to find widespread industrial applications, and enhance our understanding of smell, and in the longer-term, how it relates to other senses and language. As an initial use case of the data, we present results using machine learning-based classification of neural responses to odors as they are recorded in the mouse olfactory bulb with calcium imaging. Our larger objective is to create the equivalent of an "MNIST database for olfaction," which we call 'oMNIST,' so that researchers are able to work from a standard dataset to further the state of the art, similar to how the availability of standard datasets catalyzed research in computer vision.

1. Introduction

Smell is arguably the most primal and yet least understood of the senses. It has been key to the survival and fitness of a large number of species, for identifying or locating food, sensing danger, driving social behaviors, tracking and navigation, and much more. Smell provides vital sensory data to the brain, but remains poorly understood as a sense for a number of reasons. At the moment, we are unable to explain the relationship between the physical and perceptual properties of odors.

In contrast, human vision is relatively well understood. Objects have properties such as shape, size, and color. We tend to agree easily whether we are seeing a banana or a house, although we might initially mistake a Chihuahua for a muffin. In other words, we largely agree on what we are seeing based on shape, size, and color.

Color, which is well understood, provides an interesting reference for understanding smell. The theory of color vision is expressed in terms of *trichromacy*, *linearity*, and *opponency* — the three-dimensionality of perceptual color space, and the independence and nature of those dimensions, respectively. These concepts were formulated and experimentally validated in the late nineteenth century. As early as 1922, with the publication of the Optical Society of America's colorimetry report, clear definitions of terms were established, both psychological and physical [1]. The publication of physical standards and experimental methods allowed for the characterization of observers, culminating with the International Commission on Illumination's formalization of both colorspace and the 'standard observer' in 1931 [2].

We begin by showing why such advances were possible for color vision early on in modern science, while they remain challenging for olfaction research today. We do this by comparing the two sensory modalities at the levels of their physical stimuli, neural systems, and the nature of perceptual experiences that observers can report, as summarized in Figure 1. We argue that the complexity of olfaction, arising from properties of both stimuli and observers, demands a central role for data science and machine learning in theory development. Towards this end, we make available a standard dataset consisting of odors and the neural signatures they generate across transgenic mice. The dataset will be augmented along both dimensions – the number of odorants and their associated responses, as additional data become available. We hope to catalyze olfaction research in Neuroscience and Data Science, similar to how MNIST led to rapid advances in machine vision.

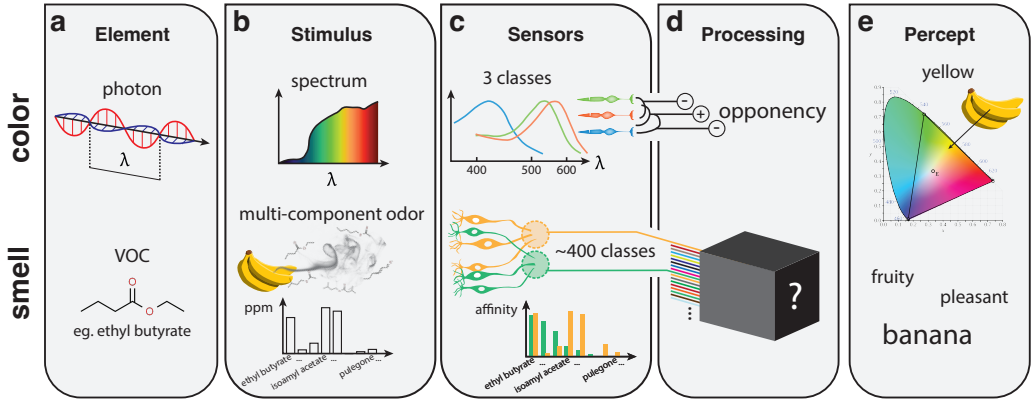

**Figure 1.** A comparison of color vision (top) and olfaction (bottom). a) The elements of the physical stimulus for color are photons, which can be defined by the single parameter of wavelength. The elements of odors are volatile organic compounds (VOCs), which have diverse molecular structures. b) Elements combine into spectra and multi-component odors. c) Just three sensor classes support trichromatic human color vision, with their well-defined spectral response profiles. In contrast humans have approximately 400 classes of olfactory receptor, with mostly undiscovered chemical receptive fields. Olfactory sensory neurons expressing the same receptor type (indicated by color) project to the same structure in the olfactory bulb, called a glomerulus. d) Human color vision is three-dimensional due to the way the three receptor type channels are compared with opponent processing. It is unclear how the information from olfactory receptor channels is compared or transformed, leaving it currently as a 'black box'. e) Color sensations can be described with semantic labels, but can also be located in the isoluminant plane, which enables accurate predictions for operations in color space such as mixing. Smells can be described in many ways, including objects of origin (banana), broader categories (fruity), pleasantness, and other descriptors, and mixing is poorly understood.

Data science aims to discover knowledge from data. The primary epistemic criterion for something to count as knowledge is its *predictive ability* [3], which is driven by the scientific philosophy that a good theory is one that makes bold predictions which survive repeated attempts at falsification [4]. Abundant data has been key to the development of strong predictive models in vision and language that underpin generative AI applications such as large language models. In contrast, olfaction data is costly to capture and hence limited. This makes it all the more important that data is collected and represented in a standard way that facilitates data pooling and theory development.

We make the case for the centrality of neural recordings to understand olfaction, and its advantages over purely psychophysical approaches. We describe a typical neural data pipeline, from acquisition to analysis, and demonstrate a worked example of predictive insights obtained through a data science approach. Like MNIST, where each digit has multiple renditions that can be used as training data, each odor in our database has multiple neural responses associated with it, along with metadata about each animal. Researchers can build on this data platform in many ways, from data handling to predictive modeling. On the data handling side, for example, we might expect innovation in methods for compressing or removing noise from the neural time series data. Indeed, we illustrate the use of two such methods and their downstream impacts on odor prediction based on simple machine learning based models. On the predictive modeling front, we demonstrate the use of AI models for prediction that aim to learn the distribution of odor space. Our longer term goal is to use these generative models to create odors from neural or linguistic descriptions.

## 2. A comparative tour through color vision and olfaction

### 2.1. Odors run through a larger gamut than colors

We begin by describing olfaction relative to color vision. Such a comparison enables us to highlight the unique aspects of olfaction and motivate our approach and the development of a standard benchmark dataset.

The physical stimulus for color vision are spectra of light, the elements of which are photons whose wavelengths fall in the visible range, namely 380-750 nm (Fig. 1.a). Each stimulus hitting the retina has a corresponding spectrum, which describes its amplitude as a function of wavelength (Fig. 1.b). While the spectrum of visible light is continuous, such that photons could have any of an infinite number of possible wavelengths, in practice wavelengths are sampled with a certain resolution. For example, if a spectrum is be reported with a wavelength resolution of 10 nm, for the visible range between 380 and 750 nm, it will be measured along 37 dimensions organized along the single linear dimension of wavelength (i.e. 380-389 nm, 390-399 nm, ...740-749 nm). A spectrum then corresponds to a point in 37-dimensional space. Spectra reaching the retina are generally determined by the emissive properties of light sources, as well as the reflective and transmissive properties of objects.

In contrast, the elements of physical stimuli for smell are volatile organic compounds, or VOCs (Fig. 1.a). Unlike photons, VOCs cannot be described by a single continuous dimension such as wavelength. Instead, the space of VOC molecular structures is discrete and highly highly multidimensional, with no obvious way of relating the myriad functional groups and moieties thaht exist in practically infinite combinations and permutations. For example, we can organize the esters ethyl propionate, ethyl butyrate, and ethyl pentanoate by the increasing carbon chain length of their parent acids, but other dimensions are required to relate branching (e.g. ethyl isovalerate) or other functional groups (e.g. ethyl benzoate). In summary, while photons can be defined by their nanometer wavelength, VOCs require the full IUPAC nomenclature, with all its prefixes, suffixes, and infixes, which indicates the complexity in naming and categorizing complex molecular structures.

The multidimensionality of VOC space presents an immediate challenge in how to select stimuli for research in olfaction. Unless a low-dimensional structure can be found in structure space, paving the way for a smaller number of 'olfactory primaries,' olfaction experiments may require the use of thousands of monomolecular odorants, which must be synthesized, purified, or purchased at significant financial and operational cost. Understanding how this large space of stimuli is sensed and how to optimize its exploration are questions that are particularly suitable for data science. At the longer term, we should understand multi-component "natural" odors that are created from the dynamic equilibria of multiple metabolic pathways, which can consist of hundreds or even thousands of VOCs (Fig. 1.b). Using more naturalistic smells may be critical for investigating natural odor abilities, for example the communication of social signals, analogous to higher-order visual processes such as the perception of faces.

### 2.2. Olfaction is a black box, not a prism

Our understanding of color vision has emerged over centuries, based a progressive understanding of light and our vision system [5, 6, 7, 8], which includes channels tuned for carrying color information as well as the concept of "opposing" colors. The discovery of three distinct and independent channels carrying color information corresponding to the three classes of cone photoreceptors in the retina (Fig. 1.c) and how they combine has been key to understanding the structure of color space.

Because of the relatively simple input-layer architecture of the human color vision system comprising three independent channels, we can liken it to a kind of coarse prism. Any incoming stimulus, whether monochromatic or made up of multiple wavelengths, gets split into three channels corresponding to the short-preferring (S), middle-preferring (M), and long-preferring (L) photoreceptor types, whose spectral response profiles are generally themselves organized along the single dimension of wavelength. The three-dimensionality of colorspace is a consequence of these three input channels, their independence, and the visual system's subsequent comparing of these channels through "opponent" processing (Fig. 1.d).

In contrast, olfactory receptors (ORs) form the largest family of G-protein coupled receptors in the mammalian genome [9], accounting for approximately 3% of the genome, with hundreds of channel types. Humans have roughly 400 receptor types, while the mouse olfactory system has roughly 1,100 types of receptors. And rather than a lock-and-key situation where ligands and receptors are exclusively matched, ORs exhibit broad tuning. A single receptor type will have affinities for a large number of odorant molecules via its promiscuous binding site, and a single odorant will bind to a large number of receptors to varying degrees. Each OR type therefore gives rise to an input channel with a broad receptive field spread throughout high-dimensional VOC structure space.

The olfaction case is further complicated by the potential for interactions between odorants at the level of receptor-ligand kinetics. While in vision photons linearly sum at the photoreceptor level, odorants may exhibit inhibitory, antagonistic and synergistic interactions when binding olfactory receptors [10]. Although it is likely such non-linear effects are more the exception than the rule, we cannot safely assume the same kind of vector space properties as Maxwell wrote to describe color, such as linear addition and interpolation to predict mixtures. Instead, the neural response is a complex time series activation of ORs.

These features, namely the large number of sensor types, the complex receptive fields of individual sensors, and the potential of interactions during olfactory signal transduction, currently render olfaction as more of a black box than the relatively transparent prism of color vision. They also prevent us from characterizing the olfactory system's input layer properties as exhaustively as we have done for color, in the form of comprehensive receptor sensitivity functions (Fig. 1.c). There are simply too many receptor types for which we would have to measure affinity for too many (practically infinite!) odorants at various levels of concentration, let alone their potential interactive combinations.

### 2.3. The diverse observer and ground truth

We complete the comparative tour of color and olfaction by considering the diversity of the population of human observers. Color vision is remarkably well-conserved among observers, who generally are in high agreement when making perceptual judgments. This is because the majority of the population are 'normal' trichromats expressing three photoreceptor types, one each from the L, M and S opsin genes. Providing participants do not have some form of congenital color vision deficiency (approx. 8% of males, 1% of females), the results from color matching experiments and other studies show minimal individual differences. This allows us to define a 'standard observer' model of how the majority of people experience color.

On the other hand, olfactory tasks often lead to disagreement among observers, in how they determine both thresholds [Variability of olfactory thresholds, Stevens (1988)] and quality [Mainland, J.D. et al. Nat. Neurosci. 17, 114–120 (2014)]. There are numerous factors that contribute to differential sensitivities of the olfactory system, including biological factors such as age, gender, and genetic ancestry, as well as factors relating to experience, lifestyle and culture. Recently it has also been shown that the particular subset of OR genes expressed, which varies wildly across individuals, also contributes to disagreements observers make about the intensity, similarity, or pleasantness of olfactory stimuli [11]. Specific anosmias, the olfactory equivalent of color vision deficiencies, where observers are insensitive to a particular odor, are widely prevalent and more of a rule than an exception [12]. For olfaction, we should therefore expect a wide distribution of diverse observers properties, with a corresponding high degree of linguistic variance associated with odor descriptions.

Despite the baked-in variance that comes with linguistic descriptors, meaningful progress is being made by relating semantic label data to the physico-chemical features of odorants. A number of models have been developed to predict semantic labels from chemical structure, with the most recent advance employing a graph neural network that takes only the molecule's atoms and chemical bonds as input to predict its linguistic descriptors taken from the Good Scents and Leffingwell & Associates food and fragrance databases, for over five thousand odorants [13]. The model was able to learn an optimized embedding function that transformed the graph of a molecule's atoms and bonds to a 256 dimensional embedding, which the authors call a principal odor map (POM). This POM is then further transformed to a read out semantic descriptors. The final trained model achieved an area under the receiver operating characteristic curve of 0.89 for a held out test set of 20% of the odors, a slight improvement over a random forest model trained to predict labels the Mordred physico-chemical descriptor dataset, which had an AUROC of 0.85 [14]. It should be noted however that both the precision (whether label predictions were correct or incorrect) and recall (whether odorant labels were predicted or missed) of this model were below 0.4, indicating the limits of language as ground truth.

While we would ultimately like to reconnect smell to perception and language, we use biology for two reasons. Fundamentally, neural responses provide an objective and robust ground truth for olfaction and may provide the foundation for a causal theory of olfaction. This is the approach we have taken towards building an olfaction database. Using a standard strain of mice, we record their neural signatures for odors multiple times and at various concentrations. Using animals from of a single genetic line limits variability at the level of OR expression. Until such time as we develop a digital nose that has an equivalent degree of sensory ability as a biological one, we will need to rely on biological sensors to provide us with the ground truth required to build predictive models.

## 3. From Odors to Odor Space - A Data Science Approach

The abundance of digital images and language content for AI has spurred research in vision and language, whereas olfaction has been hampered by its complexity and difficulty of observation. However, recent advances in monitoring neural activity using optical methods have made it easier to obtain high quality odor-evoked neural data from the olfactory bulb of the mouse. By observing responses to odors in the olfactory sensory neurons of mice *in situ*, we can understand the odorant-receptor relationships that shape the downstream olfactory process, before they are further transformed for higher processing by a diversity of observers.

Why measure responses in mice, and not directly in human observers? It would be ideal to record the activity of the human olfactory system, which would allow us to correlate perceptual judgments such as linguistic labels with their neural representations. However, current noninvasive brain recording modalities, such as fMRI, are severely limited in both temporal and spatial resolution, such that the fine-grained details of neural representations are mostly unobservable. A recent study using fMRI found odor-specific neural activity in the aorbitofrontal cortex to be predictive of linguistic descriptors [15], but the authors did not analyze activity in the olfactory bulb, the first brain region where odor information is delivered from the nose, likely due to the insufficient spatial resolution of fMRI.

Another option would be to grow cell lines, genetically modified to express human olfactory receptors, cultured to grow *in vitro*. Such cell lines can then be evaluated with ligand binding assays to measure odorant-receptor interactions. While this approach has been achieved experimentally, *in vitro* OSNs demonstrate far reduced sensitivity to those *in vivo*.

With mice, a cranial window can be surgically implanted above the olfactory bulb to gain optical access to glomeruli, the neuropil structures in the olfactory bulb where axons of olfactory sensory neurons expressing the same class of receptor conveniently aggregate (as shown in Figure 1.c). Using mouse lines which have been engineered to express genetically encoded calcium indicators, we can record odor-evoked spatial-temporal glomerular responses using a camera, seeing glomeruli literally 'light up' as a function of neural activity. Another benefit of imaging the olfactory bulb is that glomeruli for the same receptor type exhibit stereotyped spatial locations across animals, allowing for a good (but not exact) alignment of data collected in multiple mice [16]. Observing mice also enables us to build a model that is potentially transferable to humans, due to the highly conserved nature of the mammalian olfactory system with respect to both receptor subfamilies [17] and system architecture (with the exception of the vomeronasal organ and accessory olfactory bulb [18]).

A database of odor-evoked neural activity recorded from the olfactory bulb of standard mice provides a foundation for building and comparing a large variety of models in olfaction research. For example, some models may explore the relationship between the physico-chemical features of individual VOCs and their corresponding representations in the olfactory system, while others might uncover principles that govern mixing in multi-component odors, or how perceptual properties arise from olfactory stimulii. In all these cases, data that inform on stimulus-receptor relationships *in situ* within the mammalian olfactory system will provide a lot of value, as ultimately it is these relationships that determine many aspects of olfaction.

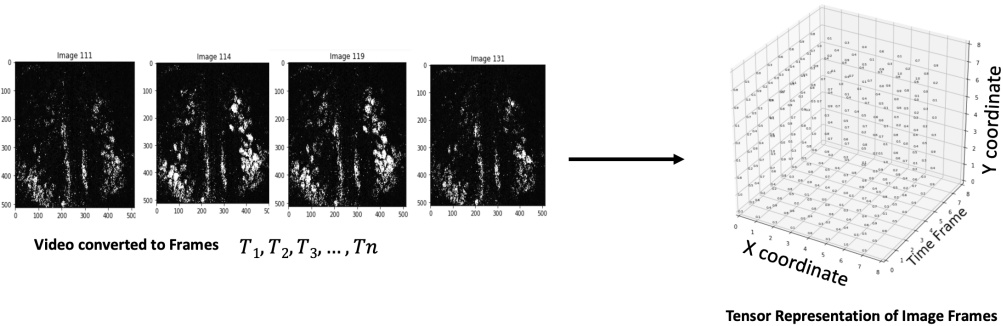

**Figure 2. Curating the Videos into Image Frames**

*The diagram above is a schematic representation of the curation of image frames from the glomeruli activation sequence video. The image stack is converted into a 3d tensor representation. The x, y and z axis represent the x, y coordinates of ROIs and image frames over the time period of the video respectively.*

In the following section, we present initial results of odor identification based on videos of glomerular activation patterns. As a proof of concept, we built an end-to-end neural data pipeline where we ingest the raw image frames of the activation videos, and remove noise from the image frames before using them to build a simple predictive model for identifying odorants from neural responses. Importantly, we show how odors map onto unique parts of the olfactory bulb. In ongoing work, we are identifying the activated glomeruli that lie in the regions of activation. Our preliminary results are a beginning for relating odorants to observable neural responses, and linking these to perception. More and better data at this level should enable us to build a causal theory of olfaction that is grounded in biology, which provides the ground truth for machine learning algorithms.

## 4. Experiments and Results
### 4.1. Dataset

The neural data were collected from mice expressing calcium indicators, GCaMP6f [19] in olfactory sensory neurons (OSNs). This biosensor, GCaMP6f is designed to fluoresce in response to changes in intracellular calcium concentrations, thus serving as an indicator for neuronal activity which is recorded using a camera.

Our database structure consists of activation patterns of two mice that are exposed to monomolecular odorants for eleven times each. For each odorant, there are twenty two images/ number of data points available for creating the train and test datasets. As an initial proof of concept, we have chosen 35 odorants listed in the Appendix that elicit strong activation of glomeruli in the dorsal olfactory bulb, and cover several important classes of odorants that include aldehydes, esters, ketones, and acids. Such diversity is essential for achieving a comprehensive perspective on glomerular activation, where we can isolate how the different chemical classes interact distinctively with specific types of olfactory receptors. At the moment, we are expanding $n$, $m$, and $x$, and making this data publicly available.

Figure 2 shows the glomeruli activation sequence for an odorant recorded as a video for each mouse. Each video file was converted to a stack of image frames in accordance with the frame rate of the camera.

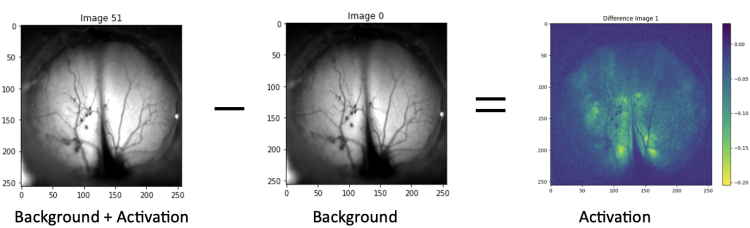

**Figure 3.** Removing Noise from Image Frames.

### 4.2. Removing Noise

The time series data is noisy. In order to extract "signal" from images, we need to remove to the extent possible, the background information that is common and dominant in the images, which shows non-functional anatomical structure such as vasculature. We are interested in the foreground pattern, which contains the potential discriminative information about glomeruli activation for each stimulus.

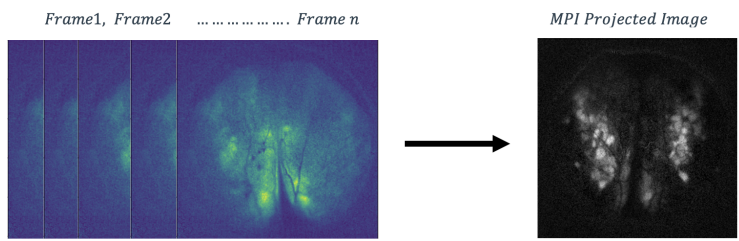

**Figure 4.** Image Frames of Glomeruli Activation Video converted to a single MPI projected image.

As a first step in removing noise, we have experimented with two commonly used algorithms, namely, simple Median Filtering [20] and Anisotropic diffusion, which is also known as Perona-Malik diffusion [21]. Figure 3 shows three image frames. The left image frame contains information about background and glomeruli activation whereas the middle image frame contains background information only. The dominance of vasculature is apparent in that it is hard to discern the difference between the leftmost image captured by the camera and the middle one, which shows the only the background. Subtracting the background image from the left image frame yields the activation shown in the rightmost part of the figure, which contains mostly the foreground activation information for the odorant.

We built a simple model by compressing the sequence of frames into a single image frame using the maximum pixel intensity of each pixel over the time series of a trial. Figure 4 is a schematic of one of the trial videos of 320 frames converted to a single MPI projected image. Each MPI projected image is processed using Anisotropic diffusion. This enhanced the image quality by removing noise and sharpen the boundaries for the ROIs as shown in 5. Each such datum is used by the machine learning algorithm to build a model that predicts the odorants based on their observed neural signatures.

### 4.3. Machine Learning

The aggregated denoised image frames were divided into train and test sets for training and evaluating a deep learning based odor-neural response model. The MPI images in training and test data correspond to different mice. Our initial goal here is to ascertain whether the results can generalize at the level of the compressed spatial representations of ROIs across subjects. If this works, we should expect models to improve considerably with more training data, where we can also consider the temporal dynamics of the neural responses across more subjects. Identifying responses at the level of glomeruli is another fertile area of research that should link specific glomeruli to odors and make the odor-response model more explicit and explainable.

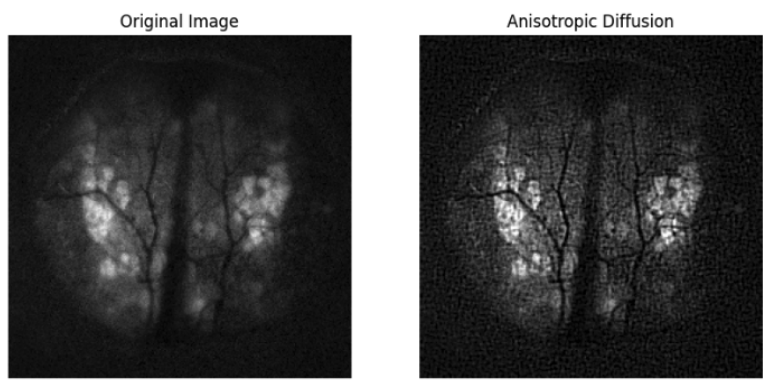

**Figure 5.** Denoising of MPIs using Anisotropic Diffusion.

We trained a convolutional neural net model based on the aggregated data. This simple architecture preserves spatial information in the images of the activated glomeruli. A description of the architecture and our model that was trained on the dataset is in Appendix B.

Table 1 shows the precision, recall and F1 score of the learned model on the test set for each odorant using the maximum likelihood approach. Figure 6 shows the confusion matrix, where each cell shows the percentage of correct responses. The AUC scores for all odorants are above 0.9. The results demonstrate unequivocally, that even a simple MPI-based model is effective at separating our odor space. We should expect the accuracy of this simple model to decline as we increase the number of odorants, and particularly odorants that elicit similarly localized responses in the dorsal olfactory bulb that we observe. Indeed, we observe a few such cases of error in our sample.

The error patterns in this rudimentary projection of time-series data reveal multiple sources of potential "noise" in the data which can lead to the prediction errors shown in the confusion matrix in Figure 6.These errors are likely due to three sources: (a) inter-subject variability, (b) similarity of the ROIs for odors, and (c) experimental variance or observational error, where things like image luminosity might vary across trials, or other types of noise such as spurious activation picked up from outside the olfactory bulb.

Figure 7a and Figure 7b shows inter-subject variability in the activation regions for Pentyl Acetate (outlined in red).

Figure 8a and Figure 8b shows the case of considerable overlap in ROIs of the MPI test images for Methyl Benzoate and Benzaldehyde (outlined in red) that result in 73% of Methyl Benzoate images being misclassified as Benzaldehyde. Although the two odorants exhibit certain structural commonalities such as their benzene ring, a characteristic that underpins their aromatic properties and potentially influences their interaction with olfactory receptors, it is crucial to acknowledge the inherent differences arising from their unique functional groups - an ester group in Methyl Benzoate and an aldehyde group in Benzaldehyde. These variations in molecular structure could lead to differential binding affinities and activation patterns across the olfactory bulb. We also observe similar overlap in the activation patterns for Propionic acid being misclassified as Valeric acid, Isobutyric acid, and Acetic acid. Similar results can be observed for Acetic Acid which is largely misclassified as Isobutyric acid, Propionic acid and Valeric acid. The molecular structures of Propionic acid and the aforementioned acids share a common carboxylic acid group (-COOH), which could be a contributing factor to their similar olfactory profiles and the resultant misclassification. This structural similarity might lead to comparable interactions with olfactory receptors, thereby triggering analogous activation patterns within the glomeruli. Additionally, odorants like Gerinol, 2 Ethylhexanal are misclassified across odorants that have different molecular structures. This intriguing misclassification underpins the intricate dynamics of the odorant-receptor interaction that determines glomerular activation.

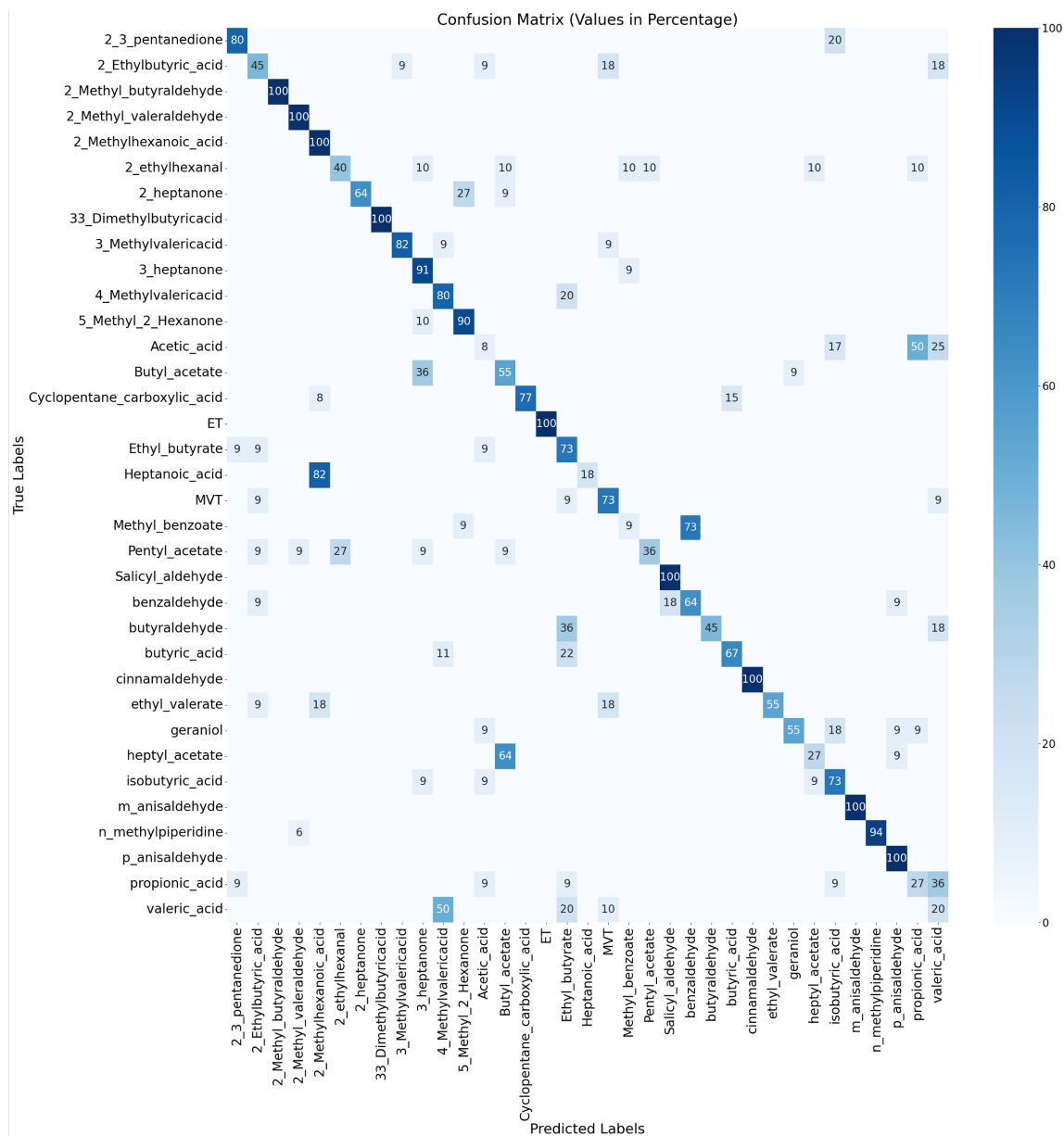

**Figure 6.** Confusion Matrix for Thirty Five Odors

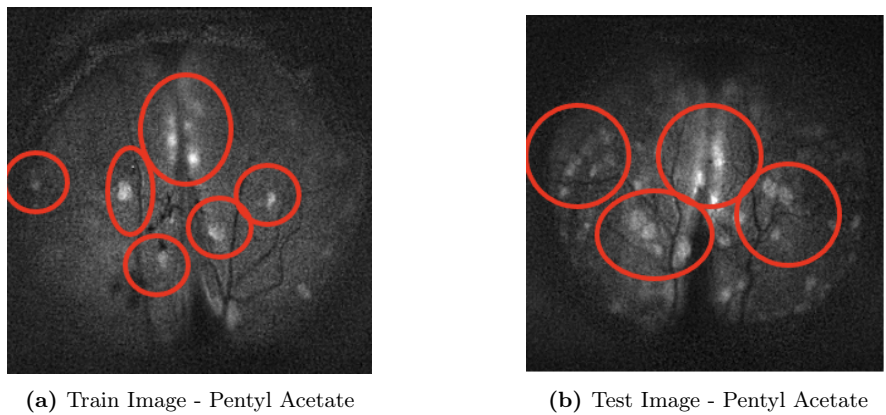

**(a)** Train Image - Pentyl Acetate          **(b)** Test Image - Pentyl Acetate

**Figure 7.** Comparison of an Instance of Train and Test Image for Pentyl Acetate
*The red outlined areas help appreciate the inter-subject variability in ROIs for Pentyl Acetate.*

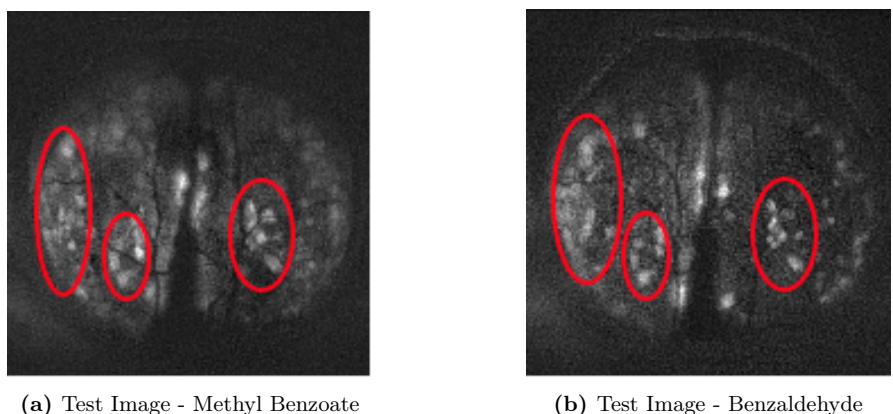

**(a)** Test Image - Methyl Benzoate          **(b)** Test Image - Benzaldehyde

**Figure 8.** Comparison of an Instance of Test Image for Benzaldehyde and Methyl Benzoate
*The red outlined areas help appreciate similarity in ROIs for Methyl Benzoate and Benzaldehyde.*

| S No. | Odorant | Precision | Recall | F1 |
|---|---|---|---|---|
| 1. | 2 3 Pentanedione | 0.80 | 0.80 | 0.80 |
| 2. | 2 Ethylbutyric Acid | 0.50 | 0.45 | 0.48 |
| 3. | 2 Methyl Butyraldehyde | 1.00 | 1.00 | 1.00 |
| 4. | 2 Methyl Valeraldehyde | 0.85 | 1.00 | 0.92 |
| 5. | 2 Methylhexanoic Acid | 0.48 | 1.00 | 0.65 |
| 6. | 2 Ethylhexanal | 0.57 | 0.40 | 0.47 |
| 7. | 2 Heptanone | 1.00 | 0.64 | 0.78 |
| 8. | 33 Dimethylbutyric Acid | 1.00 | 1.00 | 1.00 |
| 9. | 3 Methylvaleric Acid | 0.90 | 0.82 | 0.86 |
| 10. | 3 Heptanone | 0.56 | 0.91 | 0.69 |
| 11. | 4 Methylvaleric Acid | 0.53 | 0.80 | 0.64 |
| 12. | 5 Methyl 2 Hexanone | 0.45 | 0.90 | 0.60 |
| 13. | Acetic Acid | 0.17 | 0.08 | 0.11 |
| 14. | Butyl Acetate | 0.38 | 0.55 | 0.44 |
| 15. | Cyclopentane Carboxylic Acid | 1.00 | 0.77 | 0.87 |
| 16. | Ethyl Tiglate(ET) | 1.00 | 1.00 | 1.00 |
| 17. | Ethyl Butyrate | 0.40 | 0.73 | 0.52 |
| 18. | Heptanoic Acid | 1.00 | 0.18 | 0.31 |
| 19. | Methyl Valerate (MVT) | 0.57 | 0.73 | 0.64 |
| 20. | Methyl Benzoate | 0.33 | 0.09 | 0.14 |
| 21. | Pentyl Acetate | 0.80 | 0.36 | 0.50 |
| 22. | Salicyl Aldehyde | 0.83 | 1.00 | 0.91 |
| 23. | Benzaldehyde | 0.78 | 0.64 | 0.70 |
| 24. | Butyraldehyde | 1.00 | 0.45 | 0.62 |
| 25. | Butyric Acid | 0.75 | 0.67 | 0.71 |
| 26. | Cinnamaldehyde | 1.00 | 1.00 | 1.00 |
| 27. | Ethyl Valerate | 1.00 | 0.55 | 0.71 |
| 28. | Geraniol | 0.86 | 0.55 | 0.67 |
| 29. | Heptyl Acetate | 0.60 | 0.27 | 0.37 |
| 30. | Isobutyric Acid | 0.53 | 0.73 | 0.62 |
| 31. | M Anisaldehyde | 1.00 | 1.00 | 1.00 |
| 32. | N Methylpiperidine | 1.00 | 0.94 | 0.97 |
| 33. | P Anisaldehyde | 0.79 | 1.00 | 0.88 |
| 34. | Propionic Acid | 0.27 | 0.27 | 0.27 |
| 35. | Valeric Acid | 0.14 | 0.20 | 0.17 |

**Table 1.** Results on Test Data

Finally, we also see a case of spurious activation outside the bulb. To visualize this, we need to show how the trained neural network differentiates among the activation patterns for the odorants. We use a Gradient based Class Activation Mapping (Grad CAM) [22] in order to localize the features that are responsible for the prediction for each odorant. The table in Appendix A provides a visual analysis, showing the averaged MPI representations of the training and testing images alongside the grad cam-induced heat maps, which identifies the localized activations that are important for the 35 odorants. (As noted previously, the next step is to map these regions to specific glomeruli within the olfactory bulb.)

The grad cam visualizations employ a color-coded scheme to signify the model's prioritization within the bulb's regions: areas marked in red are deemed most significant by the model, followed by those in green. Blue zones are considered unimportant. Even at this aggregate level, we can see that the model can classify the odorants based on the activation patterns observed across various segments of the bulb for the odorants. Figure 9a, Figure 9b and Figure 9c show that the model considers features outside the olfactory bulb (outlined in red), which are clearly noise. This experimental error can be mitigated by pre-processing algorithms that focus exclusively on

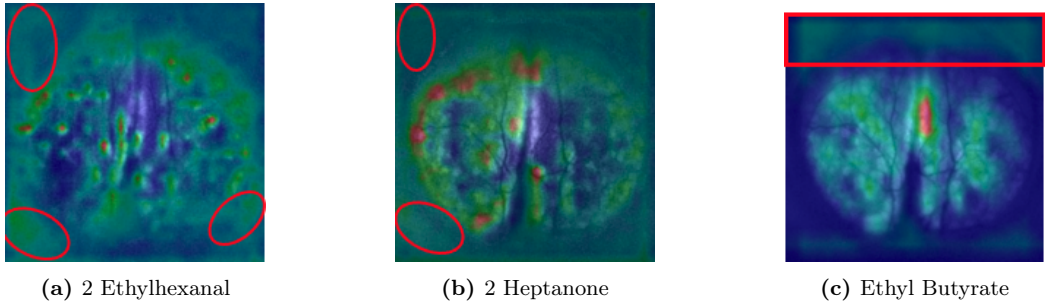

**(a)** 2 Ethylhexanal      **(b)** 2 Heptanone      **(c)** Ethyl Butyrate

**Figure 9.** GradCAM Images Show Features Learnt by Model Outside Olfactory bulb
*The Grad CAM overlay images above show the discriminative regions considered important by the model. Areas marked in red are deemed most significant by the model, followed by those in green. Blue zones are considered unimportant.*

data within the bulb. Together with denoising, such algorithms are essential in order to attenuate background interference and noise in images and achieving better discriminative acuity.

## 5. Discussion

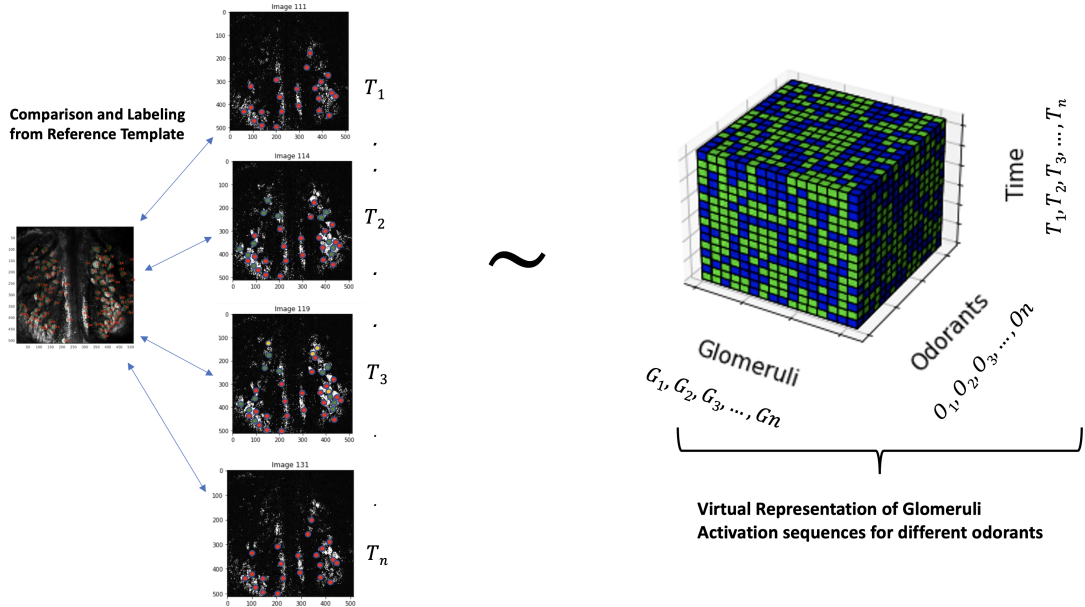

**Figure 10. A Virtual Model of Glomeruli Activation Sequence for different odorants**
*The virtual representation shows comparison and labelling the Glomeruli activation for respective frames. The cube represents the Glomeruli, Odorants and Time on three different axes.*

The Regions of Interest (ROIs) highlighted by the GradCAM reveal the features considered significant by the model. As the results indicate, odors are discriminable with this modeling approach. The next step is to identify actual glomerular regions in the images that are aligned across the subjects, The incorporation of latent biological features, such as glomerular mappings across mice, would improve both the predictive power and interpretability of such models. Another feature in the imaging data that could aid in both prediction and interpretation is time, especially as research has demonstrated the importance of timing and sequence in the perception of glomerular response patterns [23].

As a next step in this direction, we are building a model that can learn both the spatial and temporal representation of the activated glomeruli on a larger dataset of odorants. Our goal is to localize the ROIs and build a database representation as shown in Figure 10, where the $z$ axis represents odorants, the $x$ axis represents Expected Glomeruli whose activation across time is represented by the $y$ axis. This "Expected Glomeruli Activation Cube" as shown in Figure 10 can, in effect, represent the archetypal time series glomerular responses for various odorants. As we move towards a generative model of olfaction, the tensor in Figure 2 can become the output label that is to be predicted. Considering the high dimensionality and non-linearity of odor space, this will require a large database. Accordingly, we have embarked on building a thousand odor database called 'Odor-MNIST (oMNIST)', with populated data along the lines of Figures 2 and 10. We release a curated version of the data for thirty-five odorants in public domain at NYU Ultraviolet Langone Repository at https://doi.org/10.58153/j2932-aaf14 [24].

## 6. Conclusion

In summary, our position is that there is no simple way to parametrize the space of molecules into an "odor map" at the moment, and little is known about the geometry of olfaction, such as the relationship between individual molecules and mixtures that contain them. However, generative AI models such as diffusion-based deep neural networks hold considerable promise in their ability to efficiently learn the joint distribution of odor-receptor interaction.

In conclusion, Artificial Intelligence has made tremendous strides in language and vision thanks to the abundance of data and advances in algorithms that are able to learn joint distributions from such data. Smell represents an emerging frontier of AI thanks to better imaging technology for sensing this basic but complex sense, whose perception by the brain involves a combination of chemistry and signal processing. Machine learning and the tools of data science will be invaluable for solving open questions in olfaction, especially if this approach considers the latent biological variables involved, that shed light on the central role of odorant-receptor interactions in smell[25].

Accordingly, our primary goal is to provide a foundation for advancing the science of olfaction through a data science approach, which encourages the development of standard datasets against which AI models can be evaluated. Just as research in vision was catalyzed by the availability of standard datasets that were used as benchmarks in competitions, our data can be similarly used to advance the theory of olfaction.

As an initial step towards our goal, we make available a standard database of neural responses of transgenic mice on a number of odorants, which we will augment periodically. We have demonstrated the utility of our initial dataset. Until such time as we develop digital olfactory sensors, building computational models of smell that are grounded in data from chemistry, biology and neuroscience is a promising path forward in developing a well-grounded theory of olfaction and enhancing our understanding of human perception.

## 7. Acknowledgement

The authors wish to acknowledge Hiro Nakayami for collecting the data which has been used in this paper.

## Appendix A.    A Grad-CAM Representation of 35 Odors on Test Mouse

| Figure No. | Odorant Name | Avg Training Mouse Image | Avg Test Mouse Image | Grad CAM Overlay Image |
|---|---|---|---|---|
| 1. | 2 3 Pentanedione | 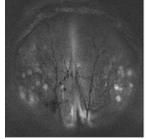 | 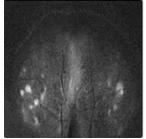 | 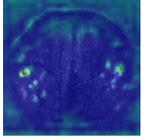 |
| 2. | 2 Ethylbutyric Acid | 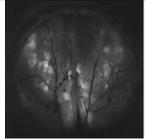 | 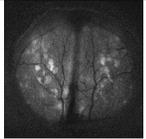 | 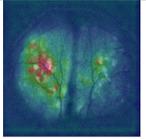 |
| 3. | 2 Methyl Butyraldehyde | 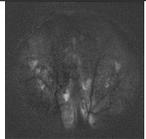 | 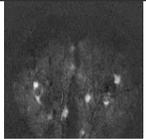 | 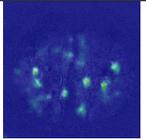 |
| 4. | 2 Methyl Valeraldehyde | 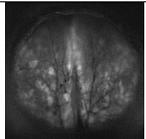 | 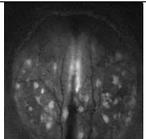 | 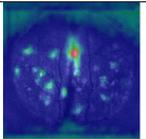 |
| 5. | 2 Methylhexanoic Acid | 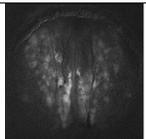 | 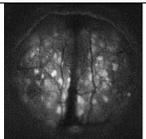 | 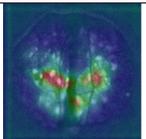 |
| 6 | 2 Ethylhexanal | 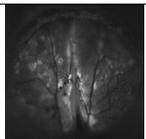 | 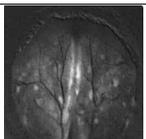 | 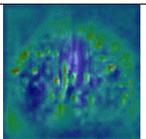 |
| 7. | 2 Heptanone | 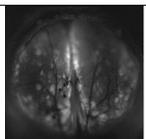 | 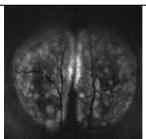 | 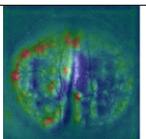 |
| 8. | 33 Dimethylbutyric Acid | 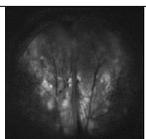 | 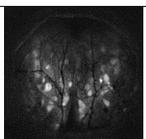 | 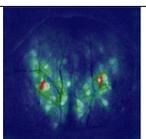 |
| 9. | 3 Methylvaleric Acid | 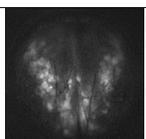 | 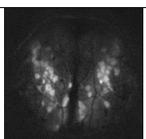 | 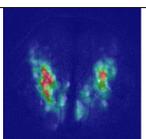 |

*The Grad CAM overlay images above show the discriminative regions considered important by the model. Areas marked in red are deemed most significant by the model, followed by those in green. Blue zones are considered unimportant. The above mapping shows the correctly classified images.*

| Figure No. | Odorant Name | Avg Training Mouse Image | Avg Test Mouse Image | Grad CAM Overlay Image |
|---|---|---|---|---|
| 10. | 3 Heptanone | 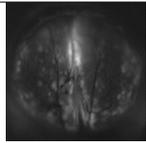 | 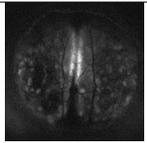 | 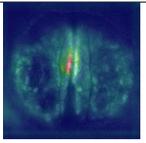 |
| 11. | 4 Methylvaleric Acid | 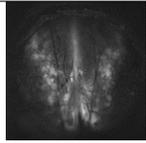 | 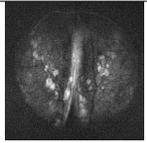 | 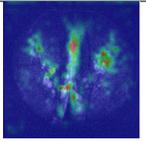 |
| 12. | 5 Methyl 2 Hexanone | 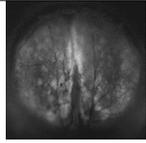 | 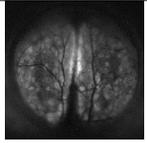 | 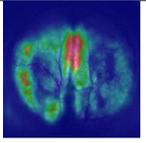 |
| 13. | Acetic Acid | 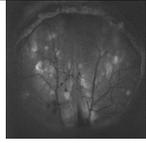 | 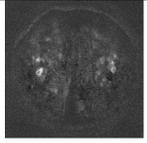 | 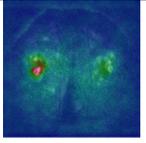 |
| 14. | Butyl Acetate | 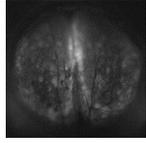 | 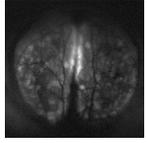 | 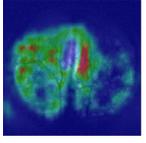 |
| 15. | Cyclopentane Carboxylic Acid | 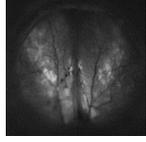 | 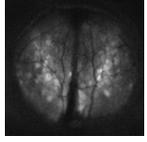 | 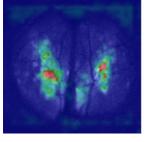 |
| 16. | Ethyl Tiglate(ET) | 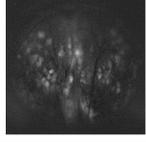 | 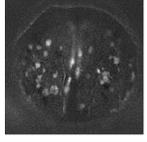 | 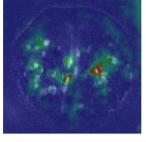 |
| 17. | Ethyl Butyrate | 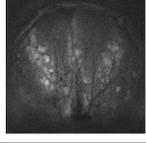 | 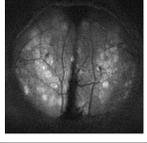 | 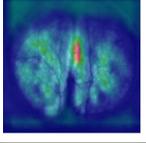 |
| 18. | Heptanoic Acid | 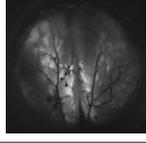 | 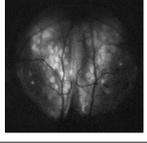 | 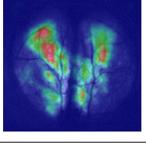 |

*The Grad CAM overlay images above show the discriminative regions considered important by the model. Areas marked in red are deemed most significant by the model, followed by those in green. Blue zones are considered unimportant. The above mapping shows the correctly classified images*

| Figure No. | Odorant Name | Avg Training Mouse Image | Avg Test Mouse Image | Grad CAM Overlay Image |
|---|---|---|---|---|
| 19. | Methyl Valerate (MVT) | | | |
| 20. | Methyl Benzoate | | | |
| 21. | Pentyl Acetate | | | |
| 22. | Salicyl Aldehyde | | | |
| 23. | Benzaldehyde | | | |
| 24. | Butyraldehyde | | | |
| 25. | Butyric Acid | | | |
| 26. | Cinnamaldehyde | | | |
| 27. | Ethyl Valerate | | | |

*The Grad CAM overlay images above show the discriminative regions considered important by the model. Areas marked in red are deemed most significant by the model, followed by those in green. Blue zones are considered unimportant. The above mapping shows the correctly classified images.*

| Figure No. | Odorant Name | Avg Training Mouse Image | Avg Test Mouse Image | Grad CAM Overlay Image |
|---|---|---|---|---|
| 28. | Geraniol | 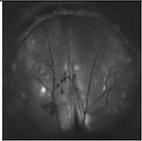 | 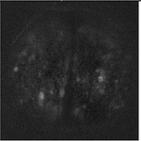 | 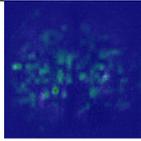 |
| 29. | Heptyl Acetate | 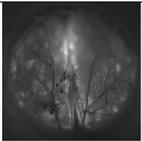 | 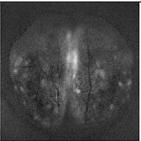 | 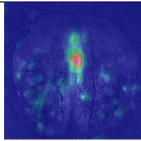 |
| 30. | Isobutyric Acid | 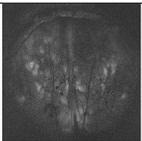 | 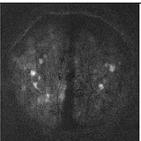 | 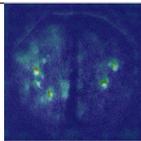 |
| 31. | M Anisaldehyde | 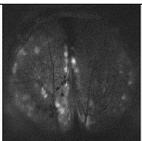 | 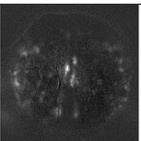 | 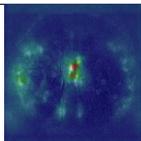 |
| 32. | N Methylpiperidine | 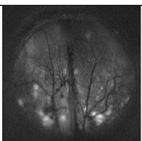 | 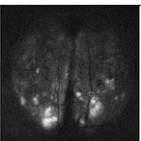 | 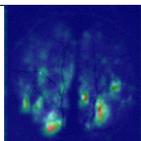 |
| 33. | P Anisaldehyde | 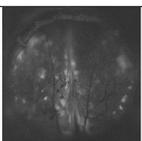 | 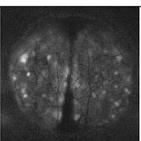 | 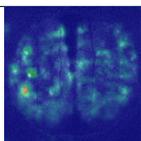 |
| 34. | Propionic Acid | 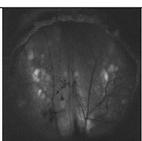 | 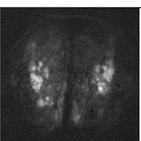 | 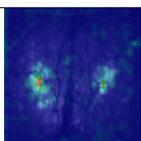 |
| 35. | Valeric Acid | 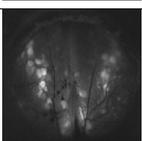 | 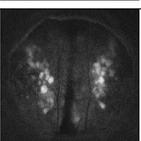 | 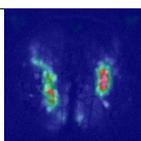 |

*The Grad CAM overlay images above show the discriminative regions considered important by the model. Areas marked in red are deemed most significant by the model, followed by those in green. Blue zones are considered unimportant. The above mapping shows the correctly classified images.*

**Appendix B.    Details of CNN Model Architecture**

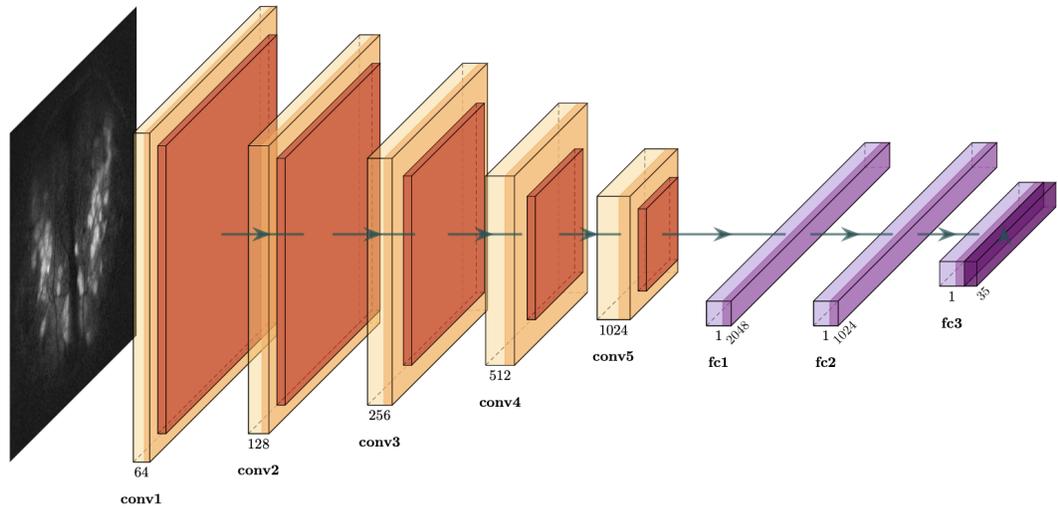

**Figure 11. Schematic Diagram of CNN Architecture**

*The diagram above is a schematic representation of the CNN used for classification of MPI images. The Layers in brown indicate the convolutional and the pooling layers. The layers in magenta are fully connected layers with a softmax layer at the end.*

The convolutional neural net used for the classification comprises five convolutional layers that take a single-channel input and applies 64 filters with a kernel size of 3x3, using a stride of 1 and padding of 1 to preserve the spatial dimensions of the input. We increase the number of filters, doubling from one layer to the next: conv2 has 128 filters, conv3 has 256 filters, conv4 has 512 filters, and conv5 has 1024 filters. Each of these layers also uses 3x3 kernels with a stride of 1 and padding of 1, enabling the network to learn increasingly complex and abstract features at each layer. Following each convolutional layer, a batch normalization layer is applied to normalize the output of the convolutional layers, reducing internal covariate shift and stabilizing the learning process. We use max pooling layer to downsample the feature maps. These high level features learned by the CNN are flattened and passed through three fully connected layers. The final fully connected layer maps the 1024-dimensional features to 35 output classes. For this multi-class classification problem, we minimize a standard cross entropy loss function and choose the class with the highest predicted probability. We used a dropout rate of 0.3 to prevent overfitting.

REFERENCES                                                                                                  20[20] Huang, Thomas S. et al. "A fast two-dimensional median filtering algorithm." IEEE Transactions on Acoustics, Speech, and Signal Processing 27 (1979): 13-18.

[21] Shapiro, Linda G.. "IEEE Transactions On Pattern Analysis And Machinee Intelligence, Vol. 12. No. 7. July 1990 Scale-Space and Edge Detection Using Anisotropic Diffusion." (2001).

[22] Selvaraju, Ramprasaath R. et al. "Grad-CAM: Visual Explanations from Deep Networks via Gradient-Based Localization." International Journal of Computer Vision 128 (2016): 336 - 359.

[23] Chong E, Moroni M, Wilson C, Shoham S, Panzeri S, Rinberg D. Manipulating synthetic optogenetic odors reveals the coding logic of olfactory perception. Science. 2020 Jun 19;368(6497):eaba2357. doi: 10.1126/science.aba2357. PMID: 32554567; PMCID: PMC8237706.

[24] Agarwal, V. K., Harvey, J. S., Nakayama, H., Dhar, V., Rinberg, D. (2024). Calcium imaging of glomeruli in the olfactory bulb of the mouse in response to thirty-five monomolecular odors [Data set]. New York University. https://doi.org/10.58153/j2932-aaf14

[25] Barwich, Ann-Sophie and Elisabeth A. Lloyd. "More than meets the AI: The possibilities and limits of machine learning in olfaction." Frontiers in Neuroscience 16 (2022): n. pag.